\documentclass[%
reprint,
superscriptaddress,
showpacs,
amsmath,amssymb,
aps,
]{revtex4-2}
\usepackage{graphicx}
\usepackage{dcolumn}
\usepackage{bm}
\usepackage{CJKutf8}
\usepackage{color}
\usepackage{amssymb}
\usepackage{amsfonts}
\usepackage{palatino}
\usepackage{esint}
\usepackage[colorlinks,urlcolor=blue,linkcolor=blue,citecolor=blue,anchorcolor=blue]{hyperref}
\makeatletter

\newcommand{\Rmnum}[1]{\expandafter\@slowromancap\romannumeral #1@}
\newcommand{\yq}[1]{\textcolor{blue}{#1}}

\makeatother
\begin{document}

\begin{CJK*}{UTF8}{gbsn}

\preprint{APS/123-QED}

\title{Stable vortex soliton arrays in disclination-fractal systems}

\author{Shuang Shen}
\affiliation{Key Laboratory for Physical Electronics and Devices, Ministry of Education, School of Electronic Science and Engineering, Xi'an Jiaotong University, Xi'an 710049, China}

\author{Milivoj R. Beli\'c}
\affiliation{College of Science and Engineering, Hamad Bin Khalifa University, 23874 Doha, Qatar}

\author{Ce Shang}
\affiliation{Aerospace Information Research Institute, Chinese Academy of Sciences, Beijing 100094, China}

\author{Yongdong Li}
\affiliation{Key Laboratory for Physical Electronics and Devices, Ministry of Education, School of Electronic Science and Engineering, Xi'an Jiaotong University, Xi'an 710049, China}

\author{Yiqi Zhang}
\email{zhangyiqi@xjtu.edu.cn}
\affiliation{Key Laboratory for Physical Electronics and Devices, Ministry of Education, School of Electronic Science and Engineering, Xi'an Jiaotong University, Xi'an 710049, China}
\affiliation{State Key Laboratory of Human-Machine Hybrid Augmented Intelligence, Institute of Artificial Intelligence and Robotics, Xi'an Jiaotong University, Xi'an 710049, China}

\date{\today}

\begin{abstract}
\noindent
Vortex light fields carry orbital angular momentum and have attracted significant attention because of their wide applications in light field manipulations, optical communications, and quantum information processing. In nonlinear media, the balance between diffraction and self-action enables the formation of vortex solitons. However, achieving the stability of vortex solitons remains challenging due to radial and azimuthal modulation instabilities. Here, we report stable vortex solitons and vortex-soliton arrays in disclination-fractal configurations with different rotational symmetries, constructed by applying disclination operations to fractal lattice structures. 
The stable vortex-soliton arrays---composed of several vortex solitons---can exist along domain walls, resulting from the disclination operation applied to the topologically trivial phase.
Their stability is verified through both linear stability analysis and direct simulations of perturbed propagation.
For comparison, vortex-soliton arrays in conventional disclination configurations are found to be completely unstable, demonstrating the significance of the fractal configuration in stabilizing these nonlinear states.
These findings reveal a new nonlinear excitation mechanism arising from the interplay between fractal geometry and disclination defects, thereby providing theoretical underpinnings for both the improved understanding of multi-field excitation phenomena in complex geometric systems and opening new avenues for the design of photonic devices based on fractal disclination structures. 
\end{abstract}

\maketitle

\end{CJK*}

\section{Introduction}
Vortices are ubiquitous natural phenomena that emerge across fluid dynamics, nonlinear optics, condensed matter physics, and astrophysics.
In optics, the optical vortex is endowed with orbital angular momentum, which is manifested in the unique helical phase front and central phase singularity~\cite{yao.aop.3.161.2011,shen.light.8.90.2019,chen.ap.7.044001.2025,lu.lma.1.202502.2026}.
Vortex beams have attracted considerable interest for a wide range of practical applications, from
information encoding to particle trapping, owing to their unique phase structure. 
Unlike vortex beams explored in free space, 
vortex beams in optical materials may experience a nonlinear response that is pronounced if the vortex beam power is sufficiently strong,
so that the interplay between light refraction and nonlinear self-action is inevitable. 
This interplay gives birth to the vortex soliton, which has a specific doughnut-shaped intensity distribution in bulk media~\cite{kivshar.po.47.291.2005} that remains invariant during propagation. 
In addition to radial modulation instability, vortex solitons are also susceptible to azimuthal modulation instability~\cite{kruglov.pla.111.401.1985,kruglov.jmo.39.2277.1992}, which constitutes an obstacle to the realization of stable vortex solitons.
Considering that the propagation characteristics, intensity distribution, and dynamical evolution of a light beam are strongly affected by the spatial configuration of the background potential, 
the use of optical lattices or waveguide arrays allows one to suppress azimuthal modulation instabilities and acquire stable vortex solitons~\cite{malomed.pre.64.026601.2001,neshev.prl.92.123903.2004,fleischer.prl.92.123904.2004,terhalle.prl.101.013903.2008,terhalle.pra.79.043821.2009}.
Optical lattices, due to their high structural tunability, rich symmetries, and ease of integration with nonlinear media, have become an ideal platform to systematically explore light-matter interactions, including the distribution of localized states and relevant nonlinear dynamical features~\cite{kartashov.rmp.83.247.2011,kartashov.nrp.1.185.2019}.
Vortex solitons in optical lattices are termed discrete vortex solitons~\cite{lederer.pr.463.1.2008,mihalache.rrp.69.403.2017,malomed.pd.399.108.2019,pryamikov.jeos.17.23.2021},
whose intensity profiles deviate from the conventional doughnut shape and are predominantly confined to individual lattice sites.

In general, the generation of vortex solitons in both bulk media and optical lattices requires an input power exceeding the critical threshold. Recently, a new kind of discrete vortex solitons formed from disclination states that are power threshold-less has been reported~\cite{huang.nano.13.3495.2024,kireev.ap.8.026013.2026}. These vortex solitons bifurcate from their linear counterparts constructed by superimposing the linear eigenstates of the disclination lattice in the topological phase.
Disclination lattices are rotational defect lattices derived from periodic waveguide lattices through sector addition or deletion operations~\cite{peterson.nature.589.376.2021,liu.nature.589.381.2021,wu.pr.9.668.2021,ren.apl.8.016101.2023,ren.light.12.194.2023}. Unlike conventional defect lattices, such as those with vacancies and dislocations, disclination lattices break the translational symmetry of periodic lattices while preserving local periodic order as their core feature, thereby enabling the construction of disclination core structures with different rotational symmetries~\cite{lin.nrp.5.483.2023}. Research indicates that optical lattices possessing distinct disclination core geometries can modify the symmetry profile of localized optical fields, consequently influencing the formation, stability, and dynamical behavior of solitons, thus offering new degrees of freedom for soliton excitation and control~\cite{zhong.rpp.2026}. Nevertheless, previous studies have been largely confined to simple periodic or quasiperiodic optical lattice systems, and systematic research on disclination effects within more complex lattice configurations remains comparatively scarce. This scarcity is addressed by the present paper.

Like disclination lattices, fractal lattices lack translational symmetry and preserve only rotational symmetry. Similarly to geometric configurations featuring self-similarity and non-integer topological dimensions, fractal structures display identical morphological features across disparate spatial scales~\cite{li.sb.67.2040.2022,zheng.sb.67.2069.2022}. In optical platforms, typical fractal-lattice structures including the Sierpiński gasket and Sierpiński carpet have been extensively explored via theoretical simulations and experimental characterizations~\cite{xu.np.15.703.2021,yang.light.9.1.2020,biesenthal.science.376.1114.2022,ren.nano.12.3829.2023,li.light.12.262.2023,zhong.light.13.2017.2024,malomed.light.14.29.2025,zhang.sb.2026,eek.prl.134.246601.2025,yan.arxiv.2026,zhang.arxiv.2026}. 
It remains an open question what physical behaviors emerge when fractal geometry is integrated with disclination defects, to construct disclination-fractal structures. This question, among others, will be explored here.

The core innovation of this work is that we place the disclination-fractal structure in the topologically trivial phase,
even though there are disclination and corner states in the topologically nontrivial phase that have been extensively studied in the past decade.
We will show that there are still special eigenmodes inside the band gap even when the system resides in a topologically trivial phase.
These eigenmodes are $4\nu$-fold degenerate and localized at domain walls along the sector boundaries induced by disclination operations, where $\nu$ denotes the order of rotational symmetry.   
Based on the degenerate eigenmodes at domain walls, vortex arrays and vortex-soliton arrays can be produced.
Note that the domain wall is distinct from the previously reported one in the disclination lattice~\cite{zhang.ol.50.5917.2025} that needs on-side detunings.
Compared with traditional single vortex solitons, vortex-soliton arrays possess richer spatial configurations and dynamical evolution characteristics, offering high-quality physical carriers and application platforms for high-capacity optical information encoding, multi-channel parallel optical manipulation, and multi-dimensional light field control. 

\begin{figure*}[t]
	\centering
	\includegraphics[width=0.65\textwidth]{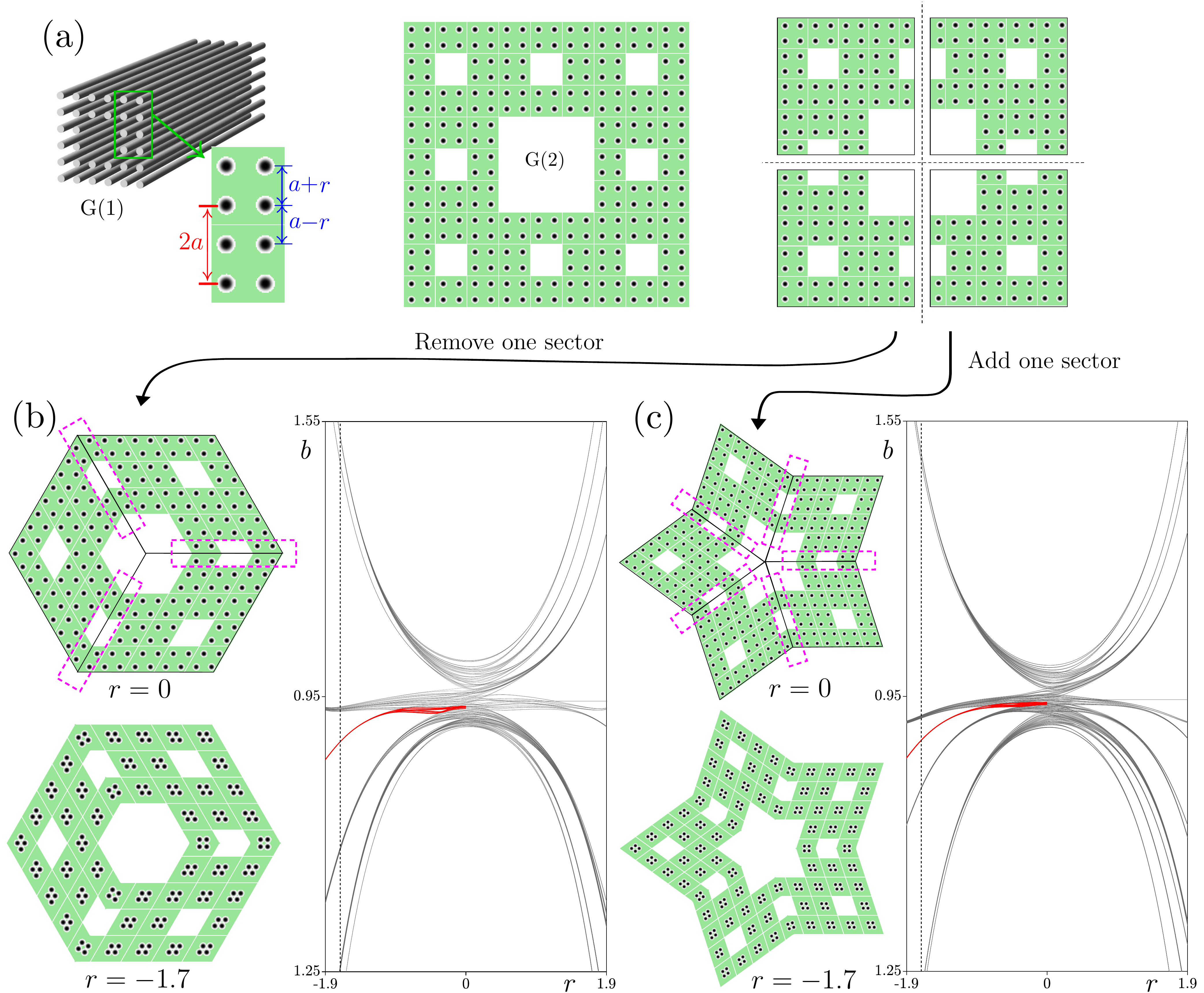}
	\caption{Disclination-fractal waveguide array and its spectrum. 
		(a) Sierpi\'nski carpets. The first-generation Sierpi\'nski carpet G(1) contains $32$ waveguides. The G($\ell$) structure is constructed by iterating G($\ell-1$), forming an array of eight G($\ell-1$) units. Based on $\mathcal{C}_4$ symmetry, this structure can be partitioned into four identical sectors. 
		(b) Removing one sector and bonding the remaining three sectors yields a disclination structure possessing $\mathcal{C}_3$ symmetry. The left panel shows structures with different shift parameter $r$. The newly formed domain walls are marked with magenta dashed rectangles. The right panel presents the linear spectrum $b$ versus $r$. 
		The red lines are the localized states on the domain wall.
		(c) Setup is as in (b) but for $\mathcal{C}_5$-symmetric disclination structure obtained by adding one sector.} 
	\label{fig01}
\end{figure*}

This work systematically investigates the existence conditions, stability properties, and dynamical propagation of vortex soliton arrays in disclination-fractal lattices. First, leveraging fractal lattice geometries and sector addition/removal operations at disclination cores, we construct disclination-fractal lattices with distinct rotational symmetries. The band dispersion and eigenmode distributions of the engineered structures are then calculated to establish vortex arrays. Second, we employ the nonlinear Schr\"odinger-like equation to characterize the propagation dynamics of vortex arrays. Numerical simulations are performed to obtain stationary vortex soliton arrays. The influence of modulation mechanisms of lattice geometry and nonlinearity strength on the formation and stability of vortex soliton arrays are comprehensively analyzed. Finally, linear stability analysis and dynamical propagation simulations are implemented to explore the evolutionary behavior of vortex-soliton arrays under external perturbations. These findings provide new physical insights into complex geometric systems and lay a theoretical foundation for the design of novel photonic devices based on disclination-fractal configurations.

\section{Theoretical model and lattices}

The propagation of paraxial beams in an optical waveguide array can be described by the paraxial wave equation~\cite{zhang.oe.18.27846.2010,jin.ap.2.046002.2020}:
\begin{align}\label{eq1}
	i\frac{\partial \mathcal{E}}{\partial \zeta} = - \frac{1}{2k}\left(\frac{\partial^2 }{\partial \xi^2}+\frac{\partial^2 }{\partial \eta^2}\right)\mathcal{E} - \frac{k}{n_0} \left( \Delta n(\xi,\eta) + n_2 |\mathcal{E}|^2 \right) \mathcal{E}.
\end{align}
Here, $\mathcal{E}$ represents the complex amplitude of the light beam, 
$\xi$ and $\eta$ represent the transverse coordinates,    
$\zeta$ is the longitudinal (propagation) coordinate, 
${k=2\pi n_0/\lambda}$ is the wavenumber with $\lambda$ being the wavelength and $n_0$ the ambient refractive index. 
$\Delta n$ is the refractive index change induced by the optical lattice
and $n_{2}$ is the nonlinear coefficient that is determined by the nonlinear optical material.

To facilitate theoretical simulations, we perform the dimensionless normalization of Eq.~(\ref{eq1}) by 
setting ${x=\xi/r_{0}}$, ${y=\eta/r_{0}}$, ${z=\zeta/kr_{0}^2}$, ${\mathcal{R}(x,y) = k^2 r_0^2 \Delta n/n_0}$, and $\psi (x,y,z) = \mathcal{E} kr_0 \sqrt{ |n_2|/n_0}$, where $r_0$ is a scaling beam width and $kr_0^2$ is the diffraction length. 
As a result, the dimensionless paraxial wave equation is:
\begin{align}\label{eq2}
	i\frac{\partial \psi}{\partial z} = -\frac{1}{2} \left(\frac{\partial^2 }{\partial x^2}+\frac{\partial^2 }{\partial y^2}\right) \psi - \mathcal{R}(x,y) \psi - |\psi|^2 \psi,
\end{align}
where $\psi$ denotes the normalized complex amplitude of the beam, $x$ and $y$ represent the transverse coordinates, and $z$ is the longitudinal coordinate.
The function $\mathcal{R}(x,y)$ is used to describe the spatial distribution of the lattice, which consists of waveguide units at individual $(x_{m,n}, y_{m,n})$ coordinates and is expressed by Gaussian functions with width $\sigma$:
\begin{align} \label{eq3}
	\mathcal{R}(x,y) = p_{\rm in} \sum_{m,n} e^{-[(x - x_{m,n})^2 + (y - y_{m,n})^2]/\sigma^2}.
\end{align}
Here, $p_{\rm in}=k^2 r_0^2 \Delta n / n_0$ represents the waveguide depth, which is proportional to the refractive index change $\Delta n$ and reflects the ability of the waveguide to confine the optical field. 
It can be seen that Eq.~(\ref{eq2}) is mathematically equivalent to the Schr\"odinger equation with cubic nonlinearity in quantum mechanics. This equation is also referred to as the optical paraxial wave Schr\"odinger equation, serving as the fundamental theoretical model for studying both linear and nonlinear propagation of light fields in waveguide arrays~\cite{szameit.jpb.43.163001.2010,longhi.lpr.3.243.2009}.
We consider only the self-focusing case, corresponding to the negative sign in the last term of Eq.~(\ref{eq2}). The continuous nonlinear Schr\"odinger equation in Eq.~(\ref{eq2}) can fully characterize the spatial  distribution of the total wave function $\psi$ inside the material, which accurately describes the coupling effects between adjacent waveguides and the dynamic evolution of modal fields within single waveguides.

In this work, the waveguide array is prepared based on the Sierpi\'nski carpet structure. As shown in Fig.~\ref{fig01}(a), the first-generation Sierpi\'nski carpet G(1) is constructed using the two-dimensional Su-Schrieffer-Heeger (SSH) model and includes 32 waveguides. 
To regulate the topological properties of the system, we fix the spacing between two unit cells at $2a$, with $a$ being the distance between nearest-neighbor waveguides, and adjust the lattice site positions by varying the parameter $r$. 
The magnified sector in the inset shows the two smallest units (indicated by green  background and 4 sites in each) with $a$ and $r$ shown clearly; all adjacent waveguides are equally spaced if ${r=0}$. 
The $\ell^{\rm th}$ generation of the Sierpi\'nski carpet structure G($\ell$) is obtained by repeating its $(\ell-1)^{\rm th}$ generation G($\ell-1$), which contains $3^{\ell+1}$ waveguides. In Fig.~\ref{fig01}(a), we illustrate the spatial structures of G(1) and G(2). 

Since the Sierpi\'nski carpet structure is $\mathcal{C}_4$ rotationally symmetric,
it can be divided into four identical sectors, as shown by the right panel in Fig.~\ref{fig01}(a).
By removing or adding a certain number of sectors and then gluing the remaining parts together, disclination-fractal structures with different rotational symmetries can be fabricated. 
As illustrated in Fig.~\ref{fig01}(b) and Fig.~\ref{fig01}(c), disclination-fractal structures with $\mathcal{C}_3$ symmetry and $\mathcal{C}_5$ symmetry are obtained, respectively. 
As marked by the dashed rectangles in Figs.~\ref{fig01}(b) and \ref{fig01}(c), this operation introduces domain walls into the disclination-fractal structures. Their favorable effects will be demonstrated in the subsequent discussion.
It is also noteworthy that self-similarity is broken in the disclination-fractal structure. Accordingly, it is impossible to derive its ``higher-order'' counterpart from the current form.
While the true fractal structure in Fig.~\ref{fig01}(a) possesses a well-established fractal dimension of ${d_f = \ln8/\ln3\sim1.893}$,
the dimensions of the structures in Figs.~\ref{fig01}(b) and \ref{fig01}(c) are not well-defined. Rather than delving into their mathematical characteristics, which are beyond the scope of this work, we focus primarily on their modal properties.

In a photonic lattice, there are intracell coupling $c_{\text{intra}}$ and intercell coupling $c_{\text{inter}}$. When the modulation parameter (e.g., the shift parameter $r$ in this work) is fixed, all intracell couplings are identical to one another across the entire lattice, as are all intercell couplings, whereas $c_{\text{intra}}$ and $c_{\text{inter}}$ generally adopt different magnitudes. Nevertheless, the cut-and-glue operation disrupts this uniform coupling configuration: it replaces the homogeneous single intracell coupling with a set of distinct intracell coupling strengths, and likewise breaks the uniformity of intercell couplings, generating multiple discrete intercell coupling values. Such variation in coupling strengths induces spectral splitting, in which an original single band splits into several separate minibands.
To suppress this spectral splitting effect and retain the uniformity of both intracell and intercell couplings, we adopt an effective manipulation technique reported in Ref.~\cite{sabour.chaos.204.117743.2026}.
This manipulation is crucial and meaningful because it indeed prevents undesirable spectral splitting and distortion in the disclination structure,
which are inevitable in previous literature~\cite{li.sb.67.2040.2022,zheng.sb.67.2069.2022,ren.apl.8.016101.2023}.

\begin{figure*}[t]
	\centering 
	\includegraphics[width=0.65\textwidth]{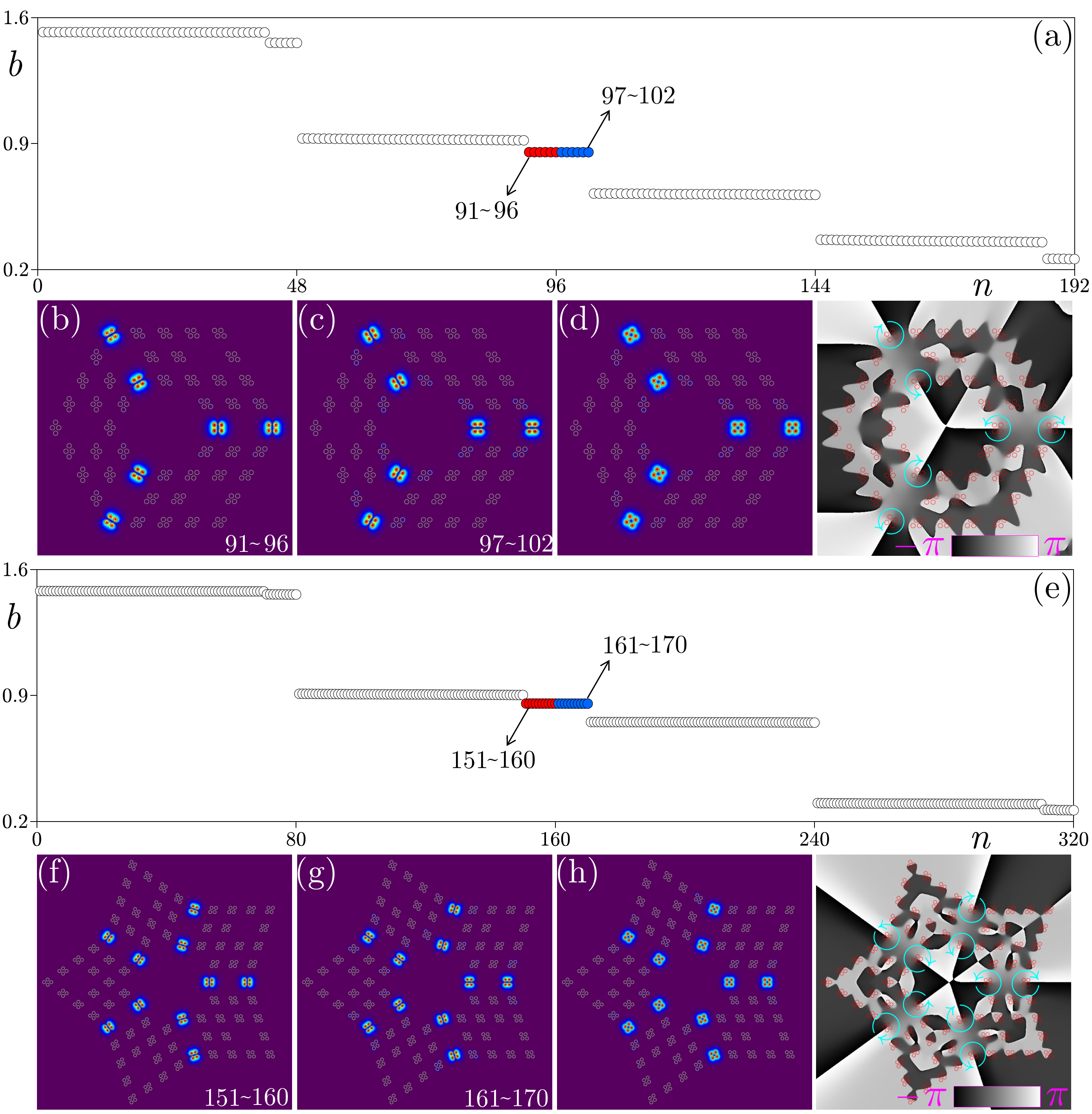}
	\caption{(a) Spectra and domain wall states with ${r = -1.7}$. 
		(b,c) Field modulus distributions of domain wall states with orthogonal polarizations. 
		(d) Field modulus of the vortex array (left panel) and its corresponding phase distribution (right panel). 
		(e-h) Setup is as (a-e) but for the results of $\mathcal{C}_5$-symmetric structure.} 
	\label{fig02}
\end{figure*}  

We would like to note that one can also obtain even higher-order fractal disclination structures from larger generation Sierpi\'nski carpets.
In general, the relation between the number $\nu$ of the vortex on each domain wall and the generation order $\ell$ of the fractal configuration is: ${\nu=2^{\ell-1}}$. As a result, one can obtain exponentially growing quantity of vortices based on higher-order generation fractal configurations.

In the following simulations, assuming that the waveguide array is fabricated in fused silica by using the femto-second laser direct-writing technique~\cite{rechtsman.nature.496.196.2013,kirsch.np.17.995.2021,arkhipova.sb.68.2017.2023,ren.light.12.194.2023,zhong.light.13.2017.2024,kompanets.am.37.2500556.2025,yan.npjn.1.40.2024}, we adopt the following typical parameters:
${r_0=10\,\mu \rm m}$, ${\lambda= 800\, \rm nm}$, ${\sigma=0.5}$, ${a=3.2}$ ($32\,\mu {\rm m}$), ${n_0=1.45}$, and ${p_{\rm in}=5}$ (${\Delta n \sim 5.6 \times 10^{-4}}$). 

\section{Results} \label{sec3}

\subsection{Domain wall states and vortex arrays}  \label{sec3a}

To investigate the spectrum of disclination-fractal lattices, we neglect the nonlinear term in Eq.~(\ref{eq2}).
To solve this linear eigenvalue problem, it is assumed that the solution of Eq.~(\ref{eq2}) is ${\psi = u(x,y) e^{ibz}}$, where $b$ is the propagation constant (an eigenvalue) and $u(x,y)$ is the field amplitude of the eigenmode. Substituting this assumed solution into Eq.~(\ref{eq2}) yields the corresponding linear eigenvalue equation:
\begin{equation}\label{eq4}
	bu = \frac{1}{2} \left( \frac{\partial^2}{\partial x^2} + \frac{\partial^2}{\partial y^2} \right) u + \mathcal{R} u.
\end{equation}
This equation can be solved by either the plane-wave expansion method or the finite-difference method. 
In Figs.~\ref{fig01}(b) and \ref{fig01}(c), we display the spectra of the $\mathcal{C}_3$- and $\mathcal{C}_5$-symmetric disclination-fractal structures, as functions of the parameter $r$.

Generally speaking, a topological phase transition occurs when a system parameter crosses a critical value~\cite{ozawa.rmp.91.015006.2019,xie.nrp.3.520.2021,zhang.nature.618.687.2023,lin.nrp.5.483.2023},
which is {$r = 0$} in the present system.
At this critical value, the inter-cell coupling equals the intra-cell coupling
and the lattice exhibits a uniform structure, as shown by the panels with ${r=0}$ in Figs.~\ref{fig01}(b) and \ref{fig01}(c).
If ${r < 0}$, the structure is in the topologically trivial phase; see the panels with ${r=-1.7}$ in Figs.~\ref{fig01}(b) and \ref{fig01}(c).
The vast majority of prior investigations mainly focus on the modes supported by systems in the topologically nontrivial phase~\cite{kirsch.np.17.995.2021,li.sb.67.2040.2022,zheng.sb.67.2069.2022}, namely structures with ${r>0}$, 
because it is commonly accepted that only extended states exist within the trivial phase. 
In the topological nontrivial regime with ${r>0}$,
there are variety of corner states, which exist at both outer corner sites and inner corner sites,
similar to those exhibited in Refs.~\cite{zheng.sb.67.2069.2022,li.sb.67.2040.2022,ren.nano.12.3829.2023,zhong.light.13.2017.2024,zhang.sb.2026}.
Thanks to the domain walls emerging in the disclination-fractal structures, for the first time we unveil the localized states under the topologically trivial phase condition.
As clearly illustrated in the spectra in Figs.~\ref{fig01}(b) and \ref{fig01}(c), the domain wall states marked in red are well isolated in wide band gaps.
In this work, we are not concerned with the states in the nontrivial phase but focus on the domain wall states in the trivial phase.

\begin{figure}[t] 
	\centering
	\includegraphics[width=\columnwidth]{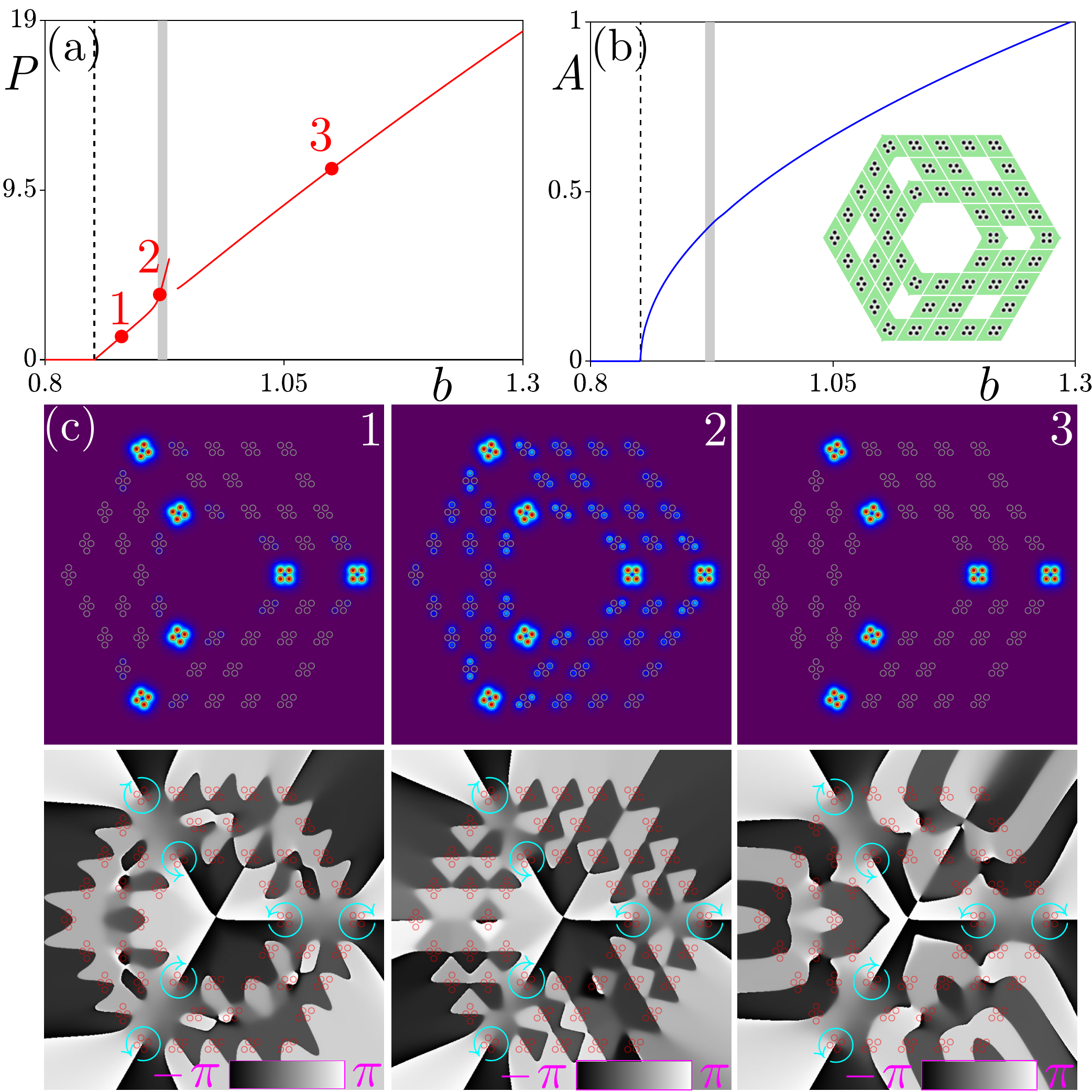}
	\caption{Family of the vortex-soliton arrays in the $\mathcal{C}_3$-symmetric disclination-fractal system with its power $P$ and peak amplitude $A$ shown in (a) and (b), respectively. The vertical dashed line indicates $b_{\rm lin}$ and the gray region includes extended states. 
		(c) Field modulus distributions (upper panels) and corresponding phases (bottom panels) of the states numbered 1, 2, and 3 in (a).
	} 
	\label{fig03}
\end{figure}

\begin{figure}[h!]
	\centering
	\includegraphics[width=\columnwidth]{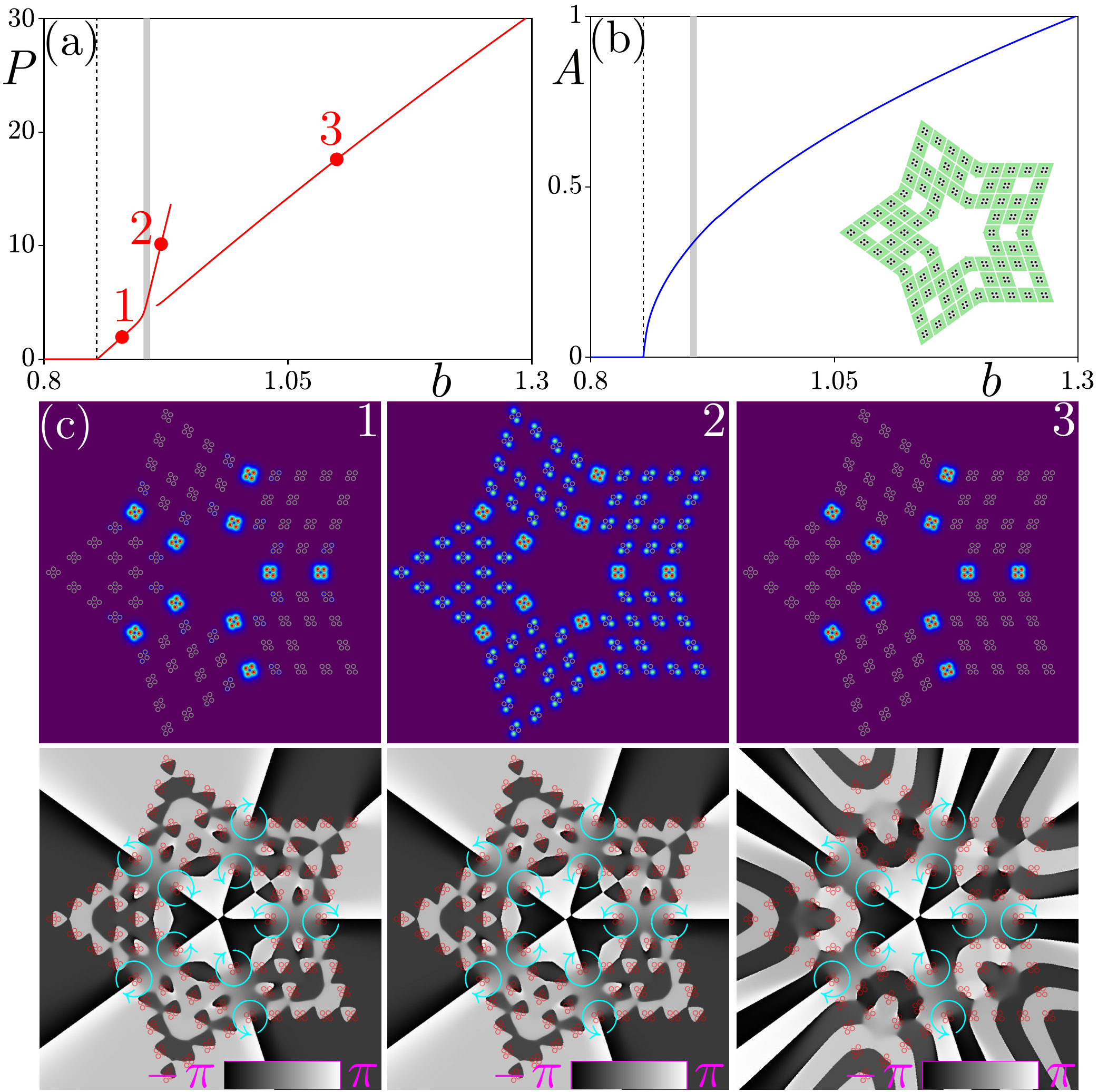}
	\caption{Setup is as Fig.~\ref{fig03}, but for the results in the $\mathcal{C}_5$-symmetric configuration.}
	\label{fig04}  
\end{figure} 

To explore the domain wall states in detail, we choose the structures with ${r=-1.7}$ as examples [indicated by the dashed lines in Figs.~\ref{fig01}(b) and \ref{fig01}(c)], and show the corresponding spectra $b$ versus mode numbers $n$ in Fig.~\ref{fig02}(a) for the $\mathcal{C}_3$-symmetric structure and in Fig.~\ref{fig02}(e) for the $\mathcal{C}_5$-symmetric structure.
As shown in Fig.~\ref{fig02}(a), there are 12 domain wall states marked in red and blue.
Even though red and blue states $u_{\rm red}$ and $u_{\rm blue}$ have different internal structure
as shown by the modulus profiles in Figs.~\ref{fig02}(b) and \ref{fig02}(c), they are all degenerate and  meet the condition ${\langle u_{\rm red}|u_{\rm blue} \rangle =0}$.
This degeneration is helpful for constructing vortex states.
We assign $u_1$ to the states numbered ${91\sim96}$ and $u_2$ to those numbered ${97\sim102}$, and the vortex state is then constructed as ${u=u_1 + i u_2}$.
According to its field modulus distribution plotted in Fig.~\ref{fig02}(d) (since this field contains multiple vortices, we refer to it as a vortex array for added accuracy), the energy of each vortex in the array is uniformly distributed over four lattice sites. 
As seen from the phase profile on the right panel of Fig.~\ref{fig02}(d), a phase singularity is located at the center of these four sites, accompanied by a phase spiral spanning across them.

For the $\mathcal{C}_5$-symmetric structure, similar results are obtained. 
As shown in Fig.~\ref{fig02}(e), the spectrum corresponding to the walls within dashed lines with ${r = -1.7}$ in Fig.~\ref{fig01}(c) has
20 degenerate red and blue domain wall states, possessing orthogonal polarizations [see modulus profiles in Figs.~\ref{fig02}(f) and \ref{fig02}(g)].
By linearly combining these degenerate domain wall states, a vortex array is constructed. The resulting field modulus profiles and the corresponding phase distributions are presented in the left and right panels of Fig.~\ref{fig02}(h). 

For the vortex array in both the $\mathcal{C}_3$- and $\mathcal{C}_5$-symmetric structures, each domain wall hosts two vortices. Their topological charges are either ${m=1}$ (marked by counterclockwise cyan circular arrows) or ${m=-1}$ (marked by clockwise cyan circular arrows). It is worth noting that the sign of the topological charge is determined by the form of the linear superposition. Substituting a minus sign for the plus sign will invert the sign of the topological charge.
Although this statement lacks mathematical rigor, assigning a topological charge of 1 to all vortices does not alter the underlying physics.
That's to say, the topological charges of vortices located on the same distance from the center can be different. Their sum is not fixed which can be $\pm1$ or $\pm3$ for the case in Fig.~\ref{fig02}(d), and $\pm1$ or $\pm5$ for the case in Fig.~\ref{fig02}(h).
In the following text, we do not mention the sign of the topological charge.

\begin{figure}[t]
	\centering
	\includegraphics[width=\columnwidth]{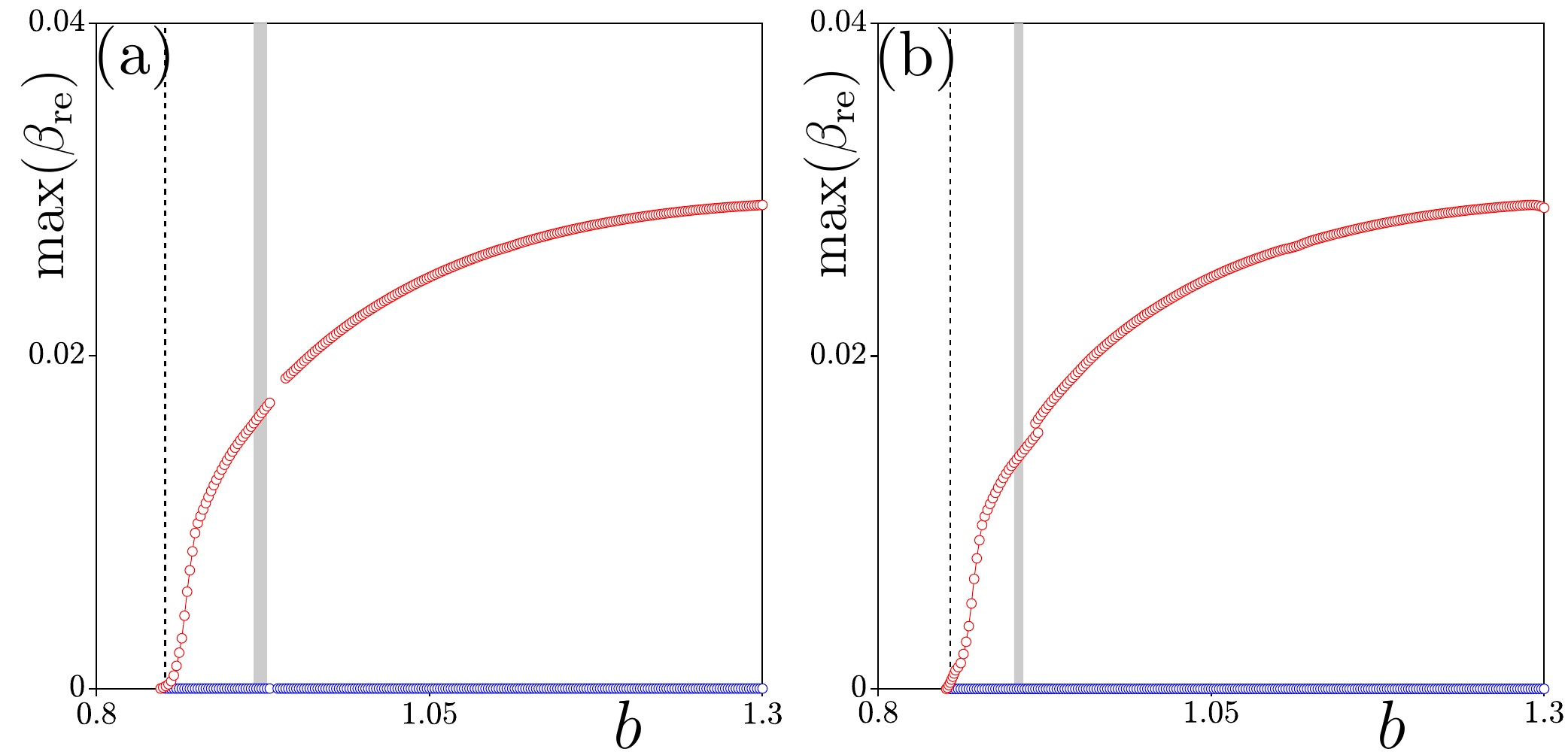}
	\caption{Maximum of the real part of the perturbation growth rate $\beta$ vs propagation constant $b$. (a) Results for the fractal disclination lattice (blue line) and the SSH disclination (red line) lattice with $\mathcal{C}_3$ symmetry. (b) Setup is as (a), but for the $\mathcal{C}_5$-symmetric case.}
	\label{fig05}  
\end{figure} 

\begin{figure*}[t]
	\centering
	\includegraphics[width=0.65\textwidth]{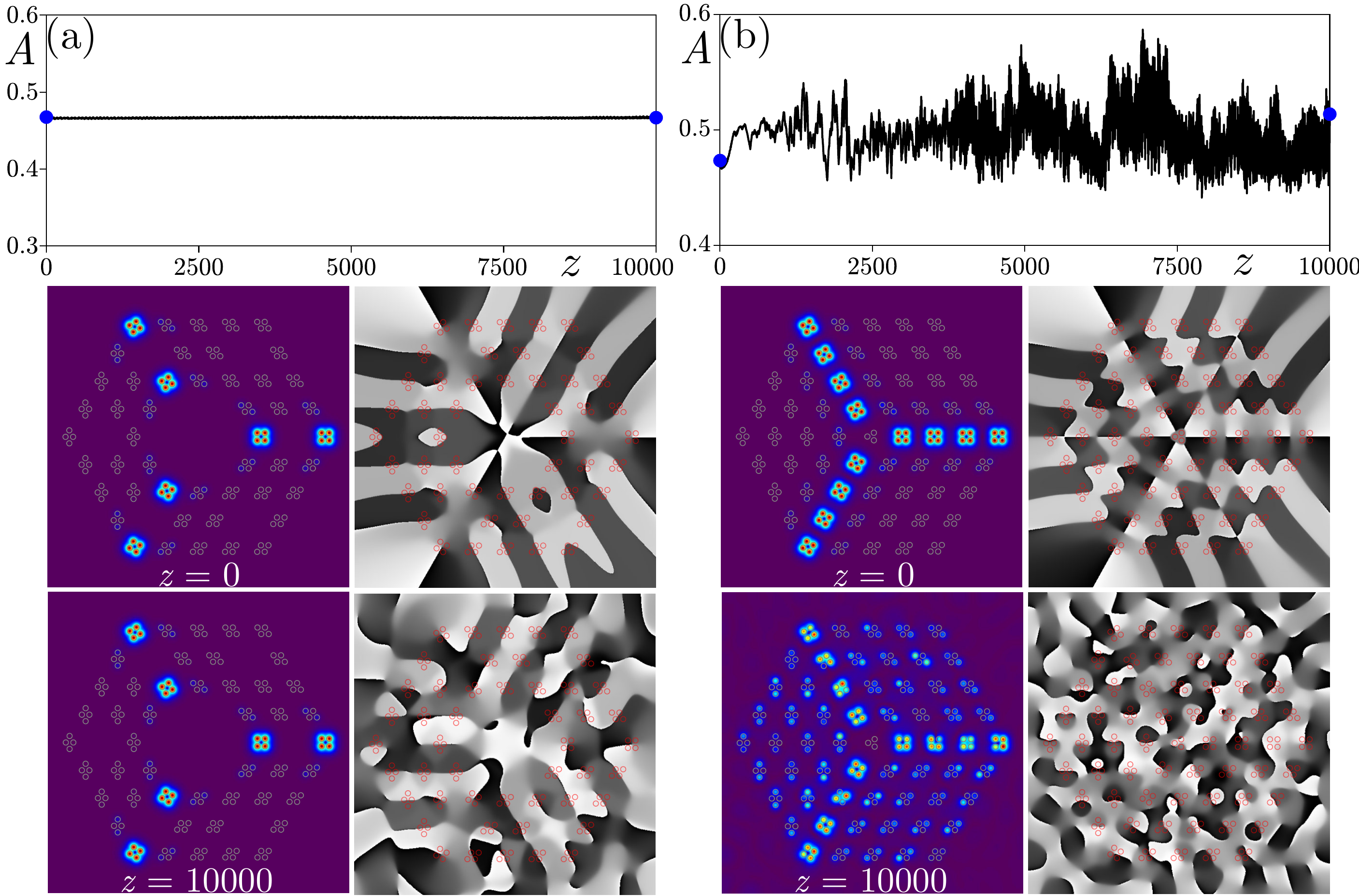}
	\caption{Propagation dynamics of perturbed vortex-soliton arrays with ${b = 0.95}$.
		(a) Stable vortex-soliton arrays in the $\mathcal{C}_3$-symmetric disclination-fractal structure. Top panel shows the evolution of the peak amplitude during propagation. Middle and bottom panels show the field mode and phase distributions at ${z=0}$ and ${z=10000}$, respectively.
		(b) Setup is as (a) but for unstable vortex-soliton arrays in corresponding disclination SSH structure. }
	\label{fig06}
\end{figure*}

\subsection{Vortex-soliton array}

One advantage of our optical platform is its convenience in exploring nonlinear phenomena. 
Based on the linear vortex array described in Sec.~\ref{sec3}\ref{sec3a}, we look for vortex-soliton arrays by including the nonlinear term of Eq.~(\ref{eq2}). With the nonlinear term neglected in Eq.~(\ref{eq4}) restored, the equation to be considered is:
\begin{equation}\label{eq5}
	bu = \frac{1}{2} \left( \frac{\partial^2}{\partial x^2} + \frac{\partial^2}{\partial y^2} \right) u + \mathcal{R}u + |u|^2 u,
\end{equation}
which can be solved using the well-known Newton iteration method.
Starting from the vortex array, one can find a family of vortex-soliton arrays with the corresponding propagation constants $b$ distributed between that of the vortex array $b_{\rm lin}$ and a predefined value larger than $b_{\rm lin}$, in this self-focusing nonlinear medium.

Figure~\ref{fig03}(a) presents the characteristic curve of vortex-soliton arrays in the $\mathcal{C}_3$-symmetric disclination-fractal system, where the horizontal axis represents the propagation constant $b$ and the vertical axis represents the total power ${P = \iint |u|^2 dxdy}$. 
The peak amplitude ${A = \max |u|}$ is shown in Fig.~\ref{fig03}(b), with the transverse landscape of the waveguide array shown in the inset.
Starting from the linear vortex array, the vortex-soliton array is obtained in the band gap between the vortex array (vertical dashed line) and the extended state band (gray region). The relation between power $P$ and the propagation constant $b$ is nearly linear, except when it is close to the extended band, where the vortex soliton array starts to carry extended state information and the slope of the curve strongly increases.
We choose two vortex-soliton arrays, numbered 1 and 2 in Fig.~\ref{fig03}(a), and display their field modulus profiles (first row) and phases (second row) in Fig.~\ref{fig03}(c).
The vortex-soliton array numbered 1 is well isolated in the band gap, so it is well localized.
Since the vortex-soliton array 2 is in the extended band, its localization is not complete and the state spills into the near-by waveguide arrays.
As the vortex-soliton array originates from its linear counterpart, we can state that the nonlinear vortex-soliton array bifurcates from the linear.

In the band gap beyond the extended state band, we find a family of vortex states with ${b \in [0.938, 1.3]}$, as shown in Fig.~\ref{fig03}(a).
No doubt, this family branch also bifurcates from the linear vortex state array; the continuity of the two families is interrupted by the extended edge band.
The vortex-soliton array numbered 3 is also shown in Fig.~\ref{fig03}(c), and exhibits a well-defined localization.
In this band gap, the degree of  localization is insensitive to variations in the propagation constant and remains unchanged. 
The peak amplitude of the vortex-soliton array in Fig.~\ref{fig03}(b) grows smoothly because the extended state information is always smaller than the peak of the vortex-soliton array.
We also find that the nonlinear effect not only counteracts the diffraction-induced broadening of the optical field, thereby preserving its localization but also controls the position of the nonlinear state within the band gap. 

Similar results are also obtained in the $\mathcal{C}_5$-symmetric lattice, as shown in Fig.~\ref{fig04}. The number of domain walls matches the order of rotational symmetry $\nu$ of the disclination-fractal lattice.
Again, benefiting from the fractal geometry, all these vortex soliton arrays maintain stability. 
In principle, one can generate the required number of stable vortex soliton arrays in disclination-fractal lattices by adding a specific amount of Frank sectors.

\renewcommand\thefigure{A\arabic{figure}}    
\setcounter{figure}{0}  

\begin{figure*}[t]
	\centering 
	\includegraphics[width=0.6\textwidth]{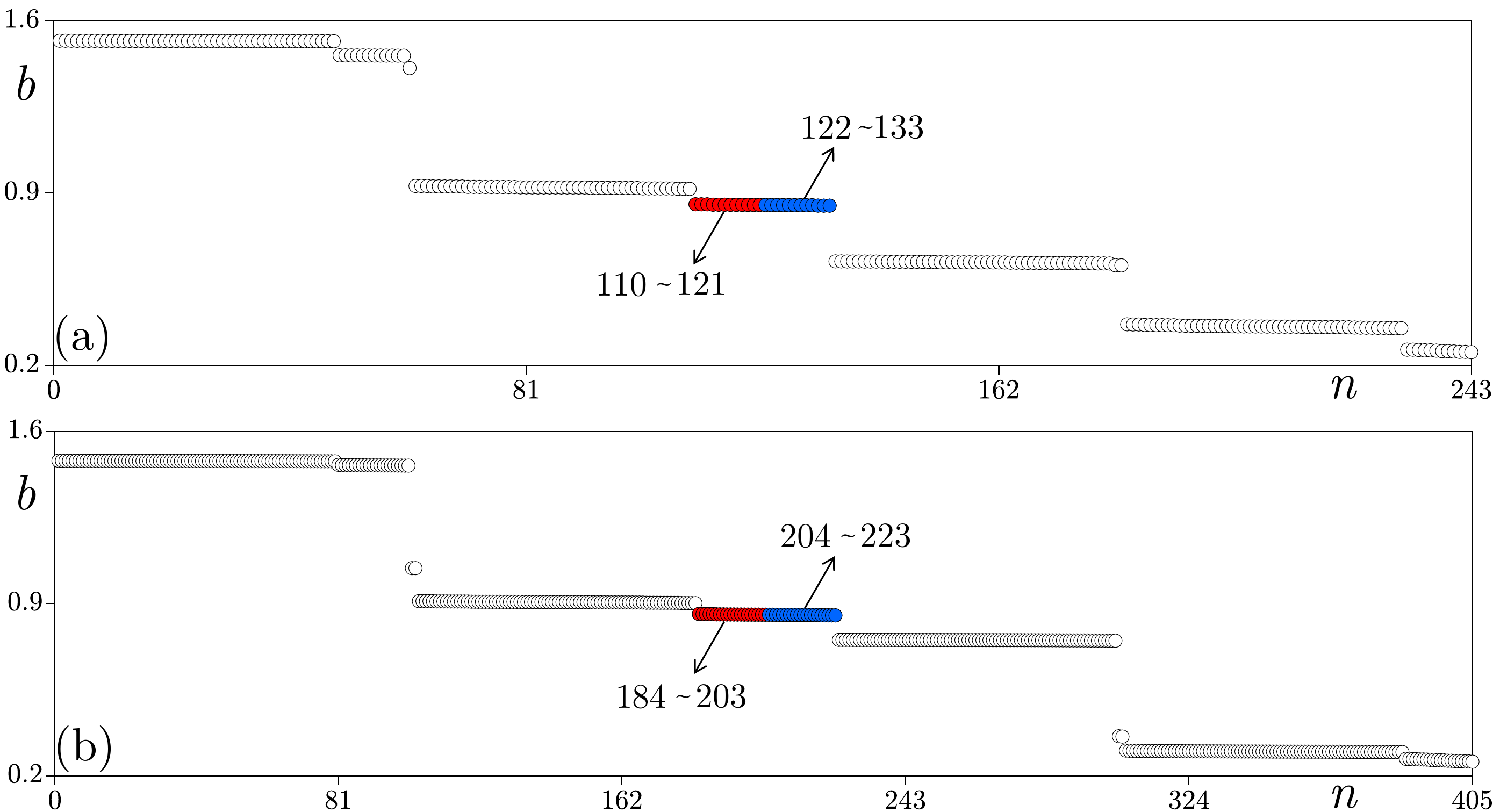}
	\caption{(a) Spectra of the $\mathcal{C}_3$-symmetric disclination structure based on the SSH model with ${r = -1.7}$. 
		(b) Setup is as in (a) but for the $\mathcal{C}_5$-symmetric structure.
		Colored dots with numbers are domain wall states.} 
	\label{fig08}
\end{figure*}  

\begin{figure}[t]
	\centering
	\includegraphics[width=\columnwidth]{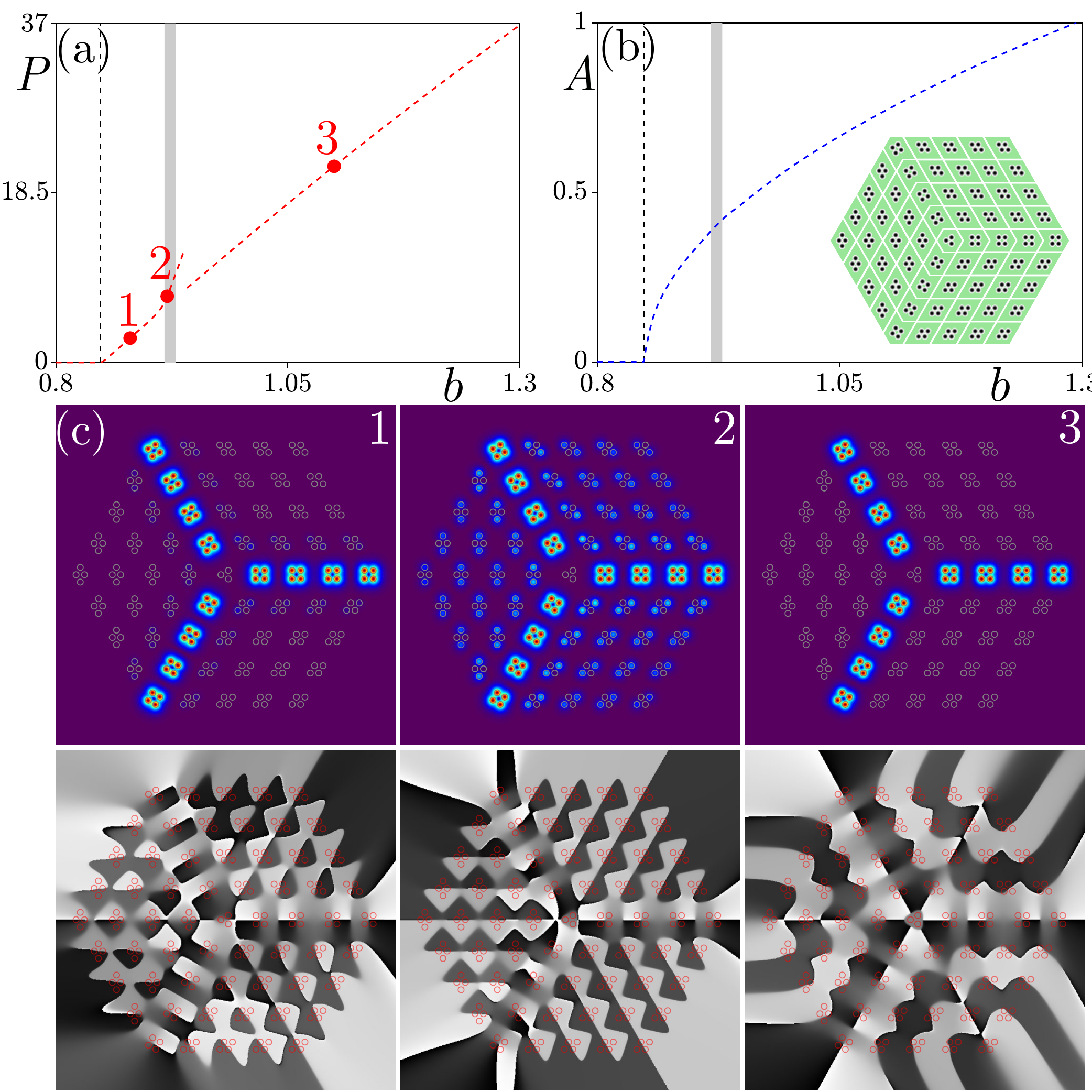}
	\caption{Setup is as Fig.~\ref{fig03}, but for the $\mathcal{C}_3$-symmetric disclination configuration based on the SSH model.}
	\label{fig03_affil}  
\end{figure} 

\begin{figure}[t]
	\centering
	\includegraphics[width=\columnwidth]{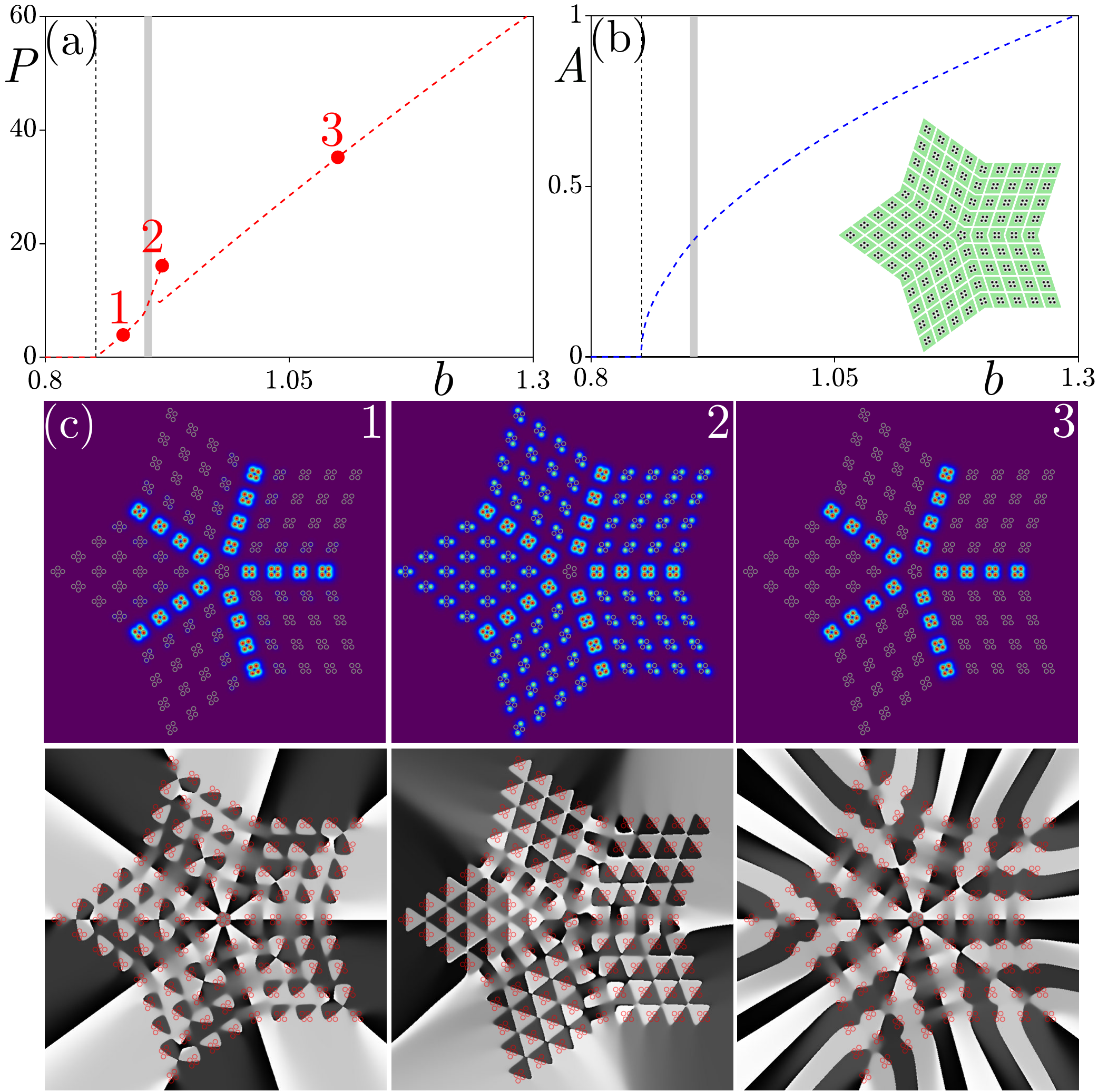}
	\caption{Setup is as Fig.~\ref{fig03_affil}, but for the $\mathcal{C}_5$-symmetric configuration.}
	\label{fig04_affil}  
\end{figure} 

\subsection{Stability analysis}

It is legitimate to wonder on the necessity of including fractal configurations in the vortex and vortex soliton arrays. The reason is simple: They offer increased stability over alternative structures. To confirm, we
present a comparative study with vortex-soliton arrays in a disclination structure based directly on the SSH model.
The spectra of the disclination configurations are displayed in Fig.~\ref{fig08} in \yq{Appendix}; the parameters are the same as those adopted in Fig.~\ref{fig02}.
The vortex-soliton array families shown in Figs.~\ref{fig03_affil} and \ref{fig04_affil} in \yq{Appendix} bifurcate from the colored dots in Fig.~\ref{fig08}.
One can see that there are also domain walls in the configuration and that vortex-soliton arrays exist on these domain walls.
In particular, the $P(b)$ dependence of the vortex-soliton array family in this system exhibits bifurcation characteristics with a physical mechanism similar to that observed in the disclination-fractal configurations. 
However, these vortex-soliton arrays are all unstable (families are indicated by dashed lines), while those in Figs.~\ref{fig03} and \ref{fig04} are all stable (families are indicated by solid lines).

To evaluate the dynamical stability of vortex-soliton arrays in the two types of disclination structure, 
we perform the linear stability analysis~\cite{ren.apl.8.016101.2023}. 
To apply this stability analysis method to our system, we introduce small perturbations $v(x,y)$ and $w(x,y)$ into the general solution of Eq.~(\ref{eq1}), as follows:
\begin{equation}
	\psi = \left[u(x,y) + v(x,y)e^{\beta z} + w^*(x,y)e^{\beta^* z}\right] e^{ibz},
\end{equation}
where $\beta$ denotes the perturbation growth rate and the asterisk represents complex conjugation. 
Substituting this expression into Eq.~(\ref{eq2}) yields a linear eigenvalue problem that determines $\beta$ for each nonlinear chiral bulk state. A state is considered unstable if the real part of $\beta$ is positive. 
In Fig.~\ref{fig05}, the blue dots represent the results in the disclination-fractal configuration, while the red dots represent those in the disclination SSH configuration.
For vortex soliton arrays in the fractal disclination lattice, the results show that the solitons exhibit excellent stability over the entire parameter range of the soliton family because the real parts of all perturbation eigenvalues are zero, which indicates that the solitons are insensitive to small perturbations. 
In contrast, vortex-soliton arrays in the disclination SSH structure exhibit distinctly different behavior---the perturbation eigenvalues in this system have positive real parts, hence the vortex soliton array exhibits exponential growth in response to small perturbations, i.e., the system is intrinsically unstable. 

\begin{figure*}[t]
	\centering 
	\includegraphics[width=0.6\textwidth]{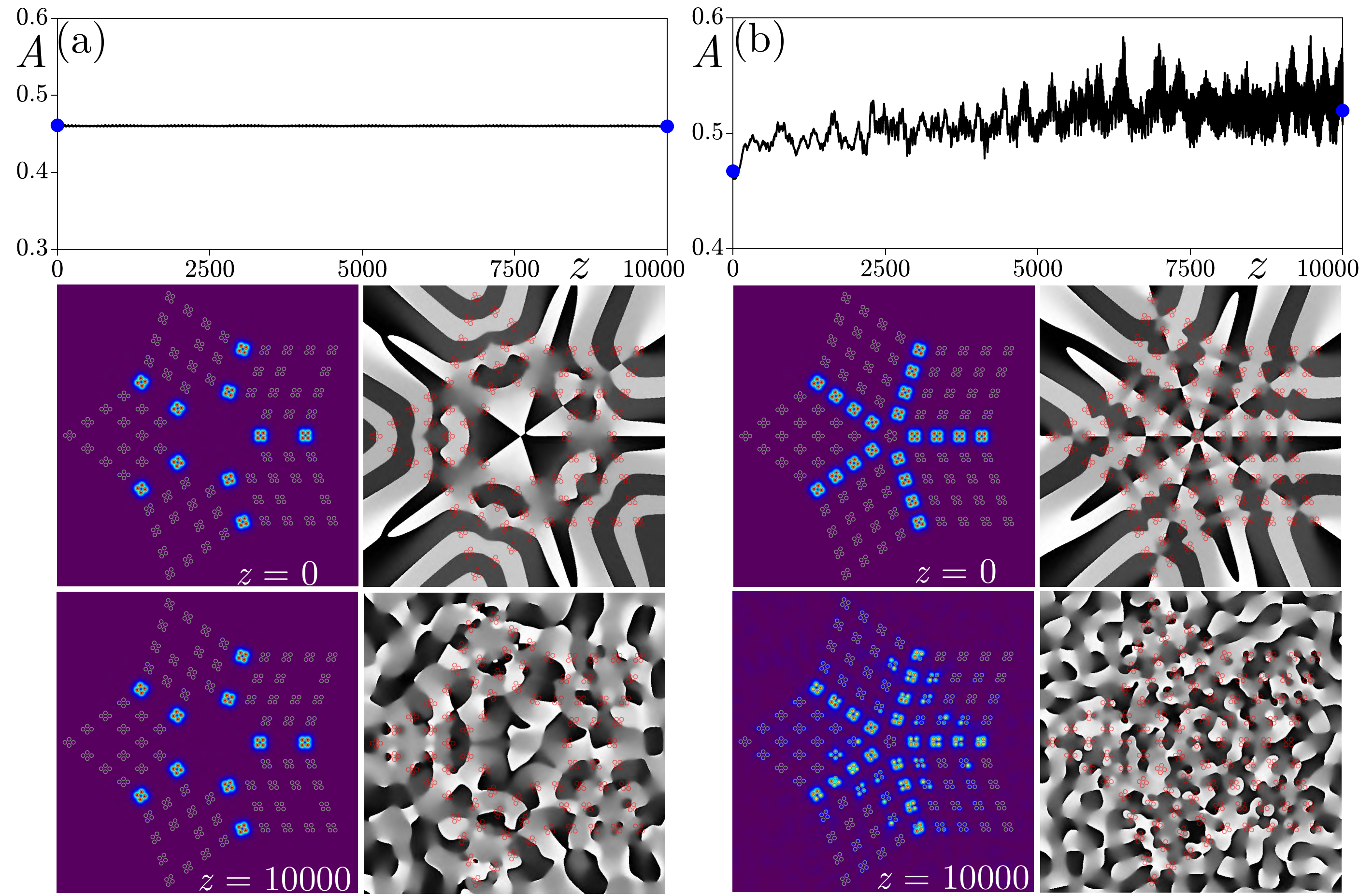}
	\caption{Setup is as Fig.~\ref{fig06} but for $\mathcal{C}_5$-symmetric structures.} 
	\label{fig07}
\end{figure*}

Furthermore, we also check the nonlinear propagation of the perturbed vortex-soliton arrays in the two different configurations.
We add a $10\%$ perturbation in the amplitude of the vortex-soliton array, and then do a propagation according to Eq.~(\ref{eq2}) using the split-step Fourier method to ${z\sim 10000}$, which is sufficiently far.
In Fig.~\ref{fig06}(a), the propagation dynamics of the perturbed vortex-soliton array in the $\mathcal{C}_3$-symmetric configuration is displayed. The results on the $\mathcal{C}_5$-symmetric configuration can be found in the \nameref{AppA} section, shown in Fig.~\ref{fig07}.

The results confirm that in the fractal disclination lattice, the vortex-soliton arrays maintain a highly consistent initial field mode distribution and exhibit a clear and stable vortex phase, even after a propagation distance as long as ${z=10000}$, with the peak amplitude remaining nearly unchanged throughout propagation.
However, for the vortex-soliton array in the disclination SSH structure, as shown in Fig.~\ref{fig06}(b), 
the stability is significantly degraded;
the vortex-soliton array gradually disintegrates during propagation, with its energy diffusing into the bulk of the structure and the phase singularity destroyed.
The agreement between the stability analysis and the perturbed propagation demonstrates that
although the two types of disclination structures are qualitatively similar in terms of the linear characteristics of soliton bifurcation, their nonlinear stability differs fundamentally. 
Due to its unique geometric configuration, the disclination-fractal lattice is more favorable for capturing stable vortex-soliton arrays. 

\section{Conclusion}
Summarizing, we have systematically investigated the nonlinear behavior of stable vortex-soliton arrays in disclination-fractal waveguide arrays in the topologically trivial phase.
By linearly combining degenerate modes in $\mathcal{C}_3$- and $\mathcal{C}_5$-symmetric configurations, we construct vortex state array, bifurcating from which the family of vortex-soliton arrays is obtained. 
Stability analysis demonstrates that these vortex-soliton arrays remain stable over their entire parameter range, confirming that the disclination-fractal lattice not only supports vortex array modes at the linear level but also ensures their nonlinear stability by virtue of its unique geometric configuration and band gap properties. 
Whereas, vortex-soliton arrays in conventional disclination structures are completely unstable.
Therefore, the disclination-fractal lattice serves as an ideal physical platform for the manipulation of stable vortex optical fields, with promising implications for applications, such as high-capacity optical storage and multidimensional light field control.  

\appendix*
\section{Other results}\label{AppA}

The spectra of $\mathcal{C}_3$- and $\mathcal{C}_5$-symmetric configurations at ${r = -1.7}$ are shown in Fig.~\ref{fig08}.
By comparing with the spectra in Fig.~\ref{fig02}, we find they are quite similar, except for the number of degenerate domain wall states.

Similar to the results shown in Figs.~\ref{fig03} and \ref{fig04}, we also display results for the corresponding disclination configurations based on the SSH model: Figs.~\ref{fig03_affil} and \ref{fig04_affil} are corresponding to the $\mathcal{C}_3$- and $\mathcal{C}_5$-symmetric structures, respectively.
Clearly, there are still domain walls in the structures and domain wall states.

Figure~\ref{fig07} displays the perturbed propagation of vortex-soliton array in $\mathcal{C}_5$-symmetric configurations, as supplementary to results shown in Fig.~\ref{fig06}.
In the disclination-fractal lattice in Fig.~\ref{fig07}(a), the vortex-soliton arrays also maintain a highly-consistent initial field mode distribution and a clear vortex phase after long-distance propagation, with the peak amplitude remaining nearly unchanged. 
In contrast, for the disclination SSH model in Fig.~\ref{fig07}(b), the vortex-soliton arrays exhibit pronounced instability under the same propagation conditions. Specifically, the field mode gradually deviates from its initial profile, the phase coherence among vortices is progressively lost, and the soliton structure eventually disintegrates. 

These results further confirm that fractal disclination lattices provide a more robust platform for stabilizing vortex soliton arrays, as compared to SSH disclination lattices, regardless of the rotational symmetry order.

\begin{acknowledgments}		
National Natural Science Foundation of China (125B2090 and 12474337); Natural Science Basic Research Program of Shaanxi Province (2024JC-JCQN-06 and 2025JCQYCX-006).
\end{acknowledgments}

%

\end{document}